V.I.Gorny, A.G.Salman, A.A.Tronin, B.V.Shilin

# TERRESTRIAL OUTGOING INFRARED RADIATION AS AN INDICATOR OF SEISMIC ACTIVITY

The analysis of satellite thermal images (STI) of the Earth's surface within the spectral range of 10.5-11.3 mcm has shown that over some linear structures of the Middle-Asian seismically active region (Kopet-Dagh, Talasso-Ferghana and other faults) there is observed a stable in time and space increasing intensity of the outgoing radiathion flux as compared to contiguous blocks (Fig.1).

A retrospective analysis of a continuous series (1980 and 1984) of observations of the outgoing IR radiation flux (daily STI taken before dawn) in the Middle-Asian region has shown that in certain individual zones of some major tectonic dislocations there appear from time to time positive anomalies of IR radiation, for instance at the point of intersection of the Talasso-Ferghana and Tamdy-Tokrauss faults. These anomalies last from 2 to 10 days. The spontaneous anomalies are characterized by a pulsating variation of area. The space confinement and duration of these anomalies permit distinguishing theme noise anomalies caused by meteorological factors. The time of the appearance of these anomalies coincides with the activation of faults over which there has been detected an increase of the outgoing IR radiation flux. In 1984 the majority of crustal earthquakes, of a magnitude over 4, in the Tien Shan were accompanied by the appearance of a positive anomaly of the IR radiation at the point of the intersection of the Tamdy-Tokrauss fault with that of Talasso-Ferghana. The area of anomalies was several tens of thousands of square kilometres.

The most outstanding example of such activization is the Ghazli earthquake of March 19, 1984 (magnitude 7.2). At the point of the intersection of the Tamdy-Tokrauss and Talasso-Ferghana faults there was detected on March 11 a positive anomaly of the outgoing IR radiation flux of exceptional intensity and enormous area (about 100 thousand sq.km) (Fig.2).

The subsequent earthquakes in the zone of the Tamdy-Tokrauss fault in the summer of 1984 (July 8, August 5,14, September 27) of the magnitudes from 4.3 to 5.3 were also preceded by the appearance of a positive anomaly of the outgoing IR radiation at the point of intersection of the Tamdy-Tokrauss and Talasso-Ferghana faults. Let us review in more detail the development in time and space of the anomalies of IR radiation at the intersection of the above faults during the time of their activation in the summer of 1984.

After the earthquake of March 19 there set in a background distribution of the terrestrial outgoing IR radiation flux (Fig.1). The appearance of a positive anomaly of IR radiation at the point of intersection of the Talaso-Ferghana and Tamdy-Tokrauss faults was noted only by the end of July 24. This anomaly developed to the south-west on June 25 and 26 along the Tamdy-Tokrauss fault. On July 27 the area of the anomaly began diminishing and on July 29 there was observed a background distribution of the terrestrial outgoing radiation flux. On July 30, August 1 and 2 there was noted the appearance of anomaly over the Tamdy-Tokrauss fault of inconsiderable area and intensity. On August 3 and 4 there was again observed in the region a background distribution of the terrestrial outgoing IR radiation flux. On August 5, 1984, there occurred a 4.3 earthquake in the area of Ghazli within the zone of the Tamdy-Tokrauss fault, with the epicentre at 40 20' N, 63 35' W.

After the earthquake, between Aug. 6 and 10 there was registered in the region a next appearance and development of the anomaly of IR radiation at the point of intersection of the same faults. The maximum area of anomaly was observed on Aug. 7 and 8. Beginning on Aug. 10, 1984, the distribution in the region of the terrestrial outgoing IR radiation flux acquires the background state. Upon this appearance of the positive anomaly of the outgoing IR radiation there followed on Aug. 14, 1984, 5.3 and 4.9 earthquakes in the Ghazli area that were genetically connected with the Tamdy-Tokrauss fault (Fig. 3)

The anomalies of the outgoing IR radiation in the zone of the Talasso-Ferghana fault appear not only prior to the earthquakes connected with the Tamdy-Tokrauss fault, but also prior to the earthquakes with epicenters in the southern spurs of the Tien Shan. The dynamics of the development in space and time of the outgoing IR radiation flux anomalies are similar to the instances cited above. It should be emphasized that the anomalies of IR radiation in the zone of the Talasso-Ferghana fault proceed only the crustal earthquakes of magnitudes over 4.3.

Similar positive anomalies of the terrestrial outgoing IR radiation flux were discovered in another region of seismic activity - in the Eastern Mediterranean. Here the anomalies of IR radiation were registered in the coastal zone (up to 300 km) on the boundary between Lybia and Egypt. These anomalies appear on the extension of the Hellenic deformation system of dislocations within the African platform, as well as over faults of the anti-Mediterranean direction in the area from the Nile delta to the Gulf of Sidra. The anomalies of IR radiation are especially distinct in this area prior to the earthquakes genetically connected with the Hellenic arch. It should be emphasized that the anomalies can appear at places considerably distant from the epicenters of the earthquakes connected with them.

The connection of the anomalies of the outgoing IR radiation over active fault zones with the time of their activation poses the problem of the nature of the appearance of such anomalies.

The relatively high velocity of the formation and development of the anomalies, their intensity reaching a few degrees, as well as the area of their development encompassing from a few up to several thousand square kilometers, deny any opportunity for connecting these anomalies directly with the processes of transformation of the mechanical energy into the thermal one in the course of maturation of an earthquake. Possible the nature of these anomalies is connected with the changes in the composition and concentration of gaseous foreign matter in the layers of atmosphere close to the earth over active faults. It is known [1-5] that active structures of the Earth's crust are characterized by high emanations of such gases as $H$, $CO$, $CO$, $CH$, $PH$, $Rn$ and other compounds. Changes in the gaseous composition of the atmosphere can result in different effects. One of these is the greenhouse effect [6]. For instance, an increase by an order (in comparison with the normal) of the content of such gases as $CO$, $CH$ in the ground layer of atmosphere 2000 m high can result in an increase of temperature at the surfase of the Earth by a few degrees, this corresponding to the intensity of the observed anomalies of the terrestrial outgoing IR radiation flux over active structures of Middle Asia and some other seismically active region. Moreover, in seismically active regions there are known facts of the increase of the concentration of these gases by 1-2 orders in the near-surface layer of atmosphere at the times of seismic activity [4,7].

It is quite possible that gases can be luminescent if the near-surface layers of atmosphere within the range of IR similarly to luminescence of the atmosphere observed in the focus zones of some earthquake [8].

O.Yu.Schmidt Institute of the Physics of the Earth
of the Academy of Sciences of the USSR, Moscow.
The All-Union Research Institute of Remote
Sensing Methods in Geology, of Satellite Airborne
Methods, Leningrad.

## Captions

Fig.1 The thermal IR image of NOAA-9 satellite. Background distribution of the terrestrial outgoing IR radiation flux at night in the Middle-Asian seismically active region (image in the spectral range of 10.5-11.3 μ, scale 1:25 000 000). Faults: I-I - Dzhalair-Naiman, II-II - Talasso-Ferghana, III-III - Tamdy-Tokrauss, IV-IV - prior to Kopet-Dagh.

Fig.2 The thermal IR image of NOAA-9 satellite on March 11, 1984 before the Ghazli earthquake of March 19, 1984 ( M=7.2 ) IR anomaly is indicated by arrows.

Fig.3 The thermal IR images of NOAA-9 satellite on: a) August 6, 1984; b) August 8, 1984; c) August 15, 1984. IR anomalies are indicated by arrows.

Fig.4 Summarized, by months, area of IR anomalies for the period of observations (top) 1979 (7 months), 1980 (4), 1984 (3), 1985 (4), 1986 (3), 1987 (5). Summarized, by months, seismic activity in Tien Shan for the same period (bottom).

УДК 528.711(202):550.34    ГЕОФИЗИКА

### В.И. ГОРНЫЙ, А.Г. САЛЬМАН, А.А. ТРОНИН, Б.В. ШИЛИН

### УХОДЯЩЕЕ ИНФРАКРАСНОЕ ИЗЛУЧЕНИЕ ЗЕМЛИ – ИНДИКАТОР СЕЙСМИЧЕСКОЙ АКТИВНОСТИ

(Представлено академиком М.А. Садовским 17 XII 1986)

Анализ тепловых космических снимков (КТС) поверхности Земли в диапазоне излучения 10,5–11,3 мкм показал, что над некоторыми линейными структурами Среднеазиатского сейсмоактивного региона (Копетдагский, Талассо-Ферганский и др. разломы) наблюдается устойчивое во времени и пространстве повышение интенсивности потока уходящего инфракрасного излучения по сравнению с сопредельными блоками (рис. 1, см. вкл. между стр. 80–81).

Ретроспективный анализ непрерывного ряда измерений потока уходящего ИК-излучения (ежесуточные КТС в предрассветное время) в Среднеазиатском регионе в 1984 и 1980 гг. показал, что в одних и тех же зонах некоторых крупных тектонических нарушений эпизодически возникают положительные аномалии ИК-излучения, например в узле пересечения Талассо-Ферганского и Тамды-Токрауского разломов. Время существования этих аномалий от 2 до 10 суток. Для эпизодических аномалий характерно пульсирующее изменение площади. Пространственная приуроченность и длительность существования этих аномалий позволяют выделять их на фоне аномалий-помех, обусловленных метеорологическими факторами. Время возникновения ИК-аномалий совпадает с активизацией разломов, над которыми зафиксировано повышение потока уходящего ИК-излучения. В 1984 г. большинство коровых землетрясений с магнитудой более 4 в зоне Тамды-Токрауского разлома сопровождались появлением положительной аномалии ИК-излучения в узле пересечения этого разлома с Талассо-Ферганским. Площадь аномалий достигала десятков тысяч квадратных километров.

Наиболее яркий пример такой активизации – Газлийское землетрясение 19 марта 1984 г. В узле пересечения Тамды-Токрауского и Талассо-Ферганского разломов 11 марта была зарегистрирована положительная аномалия потока уходящего ИК-излучения исключительной интенсивности и площади (около 100 тыс. км$^2$) (рис. 2, см. вкл. между стр. 81–82). Последовавшие затем землетрясения в зоне Тамды-Токрауского разлома летом 1984 г. (8 июля, 5, 14 августа и 27 сентября) с магнитудой от 4,3 до 5,3 также предварялись появлением положительной аномалии уходящего ИК-излучения в узле пересечения Тамды-Токрауского и Талассо-Ферганского разломов. Рассмотрим более подробно развитие во времени и пространстве аномалии ИК-излучения в узле пересечения указанных разломов в период их активизации летом 1984 г.

После землетрясения 19 марта в регионе установилось фоновое распределение потока уходящего ИК-излучения Земли (рис. 1). Лишь к концу суток 24 июля в узле пересечения Талассо-Ферганского и Тамды-Токрауского разломов отмечено появление положительной аномалии ИК-излучения. Эта аномалия развивалась в юго-западном направлении 25 и 26 июля (рис. 3) вдоль Тамды-Токрауского разлома. С 27 июля началось уменьшение площади аномалии, и 29 июля наблюдалось фоновое распределение потока уходящего излучения Земли. 30 июля, 1 и 2 августа отмечено появление незначительной по площади и интенсивности аномалии над Тамды-Токрауским разломом. 3 и 4 августа в регионе вновь наблюдалось фоновое распределение потока уходящего ИК-излучения Земли. 5 августа 1984 г. в районе г. Газли в пределах зоны Тамды-Токрауского разлома произошло землетрясение с магнитудой 4,3 и координатами эпицентра 40°20′ с.ш., 65°35′ в.д.

67

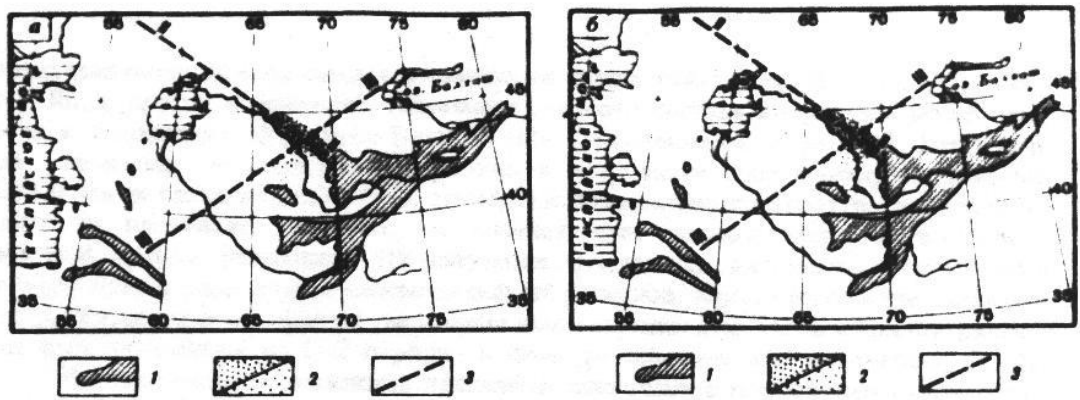

Рис. 3. Схемы дешифрирования КТС от 25 VII (*а*) и 26 VII 1984 (*б*). *1* — горные системы; *2* — аномалия потока ИК-излучения; *3* — разломы: Талассо-Ферганский (*II–II*), Тамды-Токраусский (*III–III*)

После землетрясения в регионе с 6 по 10 августа 1984 г. зарегистрировано вторичное появление и развитие аномалии ИК-излучения в узле пересечения тех же разломов. Максимальная площадь аномалии наблюдалась 7 и 8 августа. С 10 августа 1984 г. распределение потока уходящего ИК-излучения Земли принимает в регионе фоновое состояние. За вторичным появлением положительной аномалии уходящего ИК-излучения 14 августа 1984 г. последовали землетрясения с магнитудой 5,3 и 4,9 в районе Газли, генетически связанные с Тамды-Токраусским разломом.

Аномалии уходящего ИК-излучения в зоне Талассо-Ферганского разлома появляются не только перед землетрясениями, связанными с Тамды-Токраусским разломом, но и перед землетрясениями с эпицентрами в южных отрогах Чаткальского хребта. Динамика развития в пространстве и времени аномалий уходящего потока ИК-излучения аналогична описанным выше примерам. Необходимо подчеркнуть, что аномалии ИК-излучения в зоне Талассо-Ферганского разлома предшествуют только коровым землетрясениям с магнитудой более 4,3.

Аналогичные положительные аномалии потока уходящего ИК-излучения Земли обнаружены в другом сейсмоактивном регионе — Восточном Средиземноморье. Здесь аномалии ИК-излучения зарегистрированы в прибрежной зоне (до 300 км) на границе Ливии и Египта. Эти аномалии появляются на продолжении Эллинской деформационной системы дислокаций в пределах Африканской платформы, а также на разломах антисредиземноморского направления на участке от дельты Нила до залива Сидр. Наиболее ярко на этом участке аномалии ИК-излучения проявляются перед землетрясениями, генетически связанными с Эллинской дугой. Необходимо подчеркнуть значительную удаленность места появления аномалий от эпицентров связанных с ними землетрясений.

Связь аномалий потока уходящего ИК-излучения над активными разломными зонами с периодом их активизации ставит вопрос о природе возникновения таких аномалий.

Относительно высокая скорость формирования и развития аномалий, их интенсивность порядка нескольких градусов, а также площадь развития в несколько и более тысяч квадратных километров отвергают возможность прямо связать эти аномалии с процессами преобразования механической энергии в тепловую при подготовке землетрясений. Возможно, что природа этих аномалий связана с изменением состава и концентрации газовых примесей в приземных слоях атмосферы над активными разломами. Известно [1–5], что активные структуры земной коры

68

характеризуются повышенными эманациями таких газов, как $H_2$, CO, $CO_2$, $CH_4$, $PH_3$, Rn, и других соединений. Изменение газового состава атмосферы может привести к различным эффектам. Один из них - парниковый эффект [6]. Например, при увеличении на порядок содержания в приземном слое атмосферы высотой 2000 м таких газов, как $CO_2$, $CH_4$, приведет к повышению температуры поверхности Земли на несколько градусов, что соответствует интенсивности обнаруженных аномалий потока уходящего ИК-излучения Земли над активными структурами Средней Азии и ряда других сейсмоактивных регионов. Кроме того, в сейсмоактивных регионах известны факты увеличения концентрации этих газов в приповерхностном слое атмосферы на 1-2 порядка в периоды сейсмической активности [4, 7].

Не исключена возможность люминесценции газов в приповерхностных слоях атмосферы в ИК-диапазоне аналогично свечению атмосферы, наблюдаемому в очаговых зонах некоторых землетрясений [8].

Институт физики Земли им. О.Ю. Шмидта
Академии наук СССР, Москва
Всесоюзный научно-исследовательский институт
космо-аэрометодов
Ленинград

УДК 551.14                                                                                                         ГЕОФИЗИКА

Р.Ю. КУКУЛИЕВА, М.Г. КОГАН

ИЗГИБНАЯ ЖЕСТКОСТЬ ЛИТОСФЕРЫ ЕВРАЗИИ

*(Представлено академиком М.А. Садовским 12 II 1987)*

В настоящее время большинство исследователей считают, что горные пояса на континентах образуются при столкновении литосферных плит [1]. Большие топографические высоты возникают благодаря таким процессам, как надвиги по разломам и пододвигание одной плиты под другую, поскольку относительно легкая континентальная кора сопротивляется погружению на значительные глубины. Изучая геологию и геофизику древних и современных горных поясов, мы продвигаемся в понимании динамических процессов в мантии и термомеханического поведения литосферы.

Изостатическая компенсация континентальных горных поясов исследовалась рядом авторов ([2-5] и др.). Для изучения изостазии континентов часто приме-